# Bipolar Magnetic Semiconductors: A New Class of Spintronics Materials

Xingxing Li [1], Xiaojun Wu [1,2], Zhenyu Li [1], Jinlong Yang [1,*], and J. G. Hou [1]

[1] Hefei National Laboratory of Physical Science at the Microscale, University of Science and Technology of China, Hefei, Anhui 230026, China. [2] CAS Key Laboratory of Materials for Energy Conversion and Department of Materials Science and Engineering, University of Science and Technology of China, Hefei, Anhui 230026, China.

Electrical control of spin polarization is very desirable in spintronics, since electric field can be easily applied locally in contrast with magnetic field. Here, we propose a new concept of bipolar magnetic semiconductor (BMS) in which completely spin-polarized currents with reversible spin polarization can be created and controlled simply by applying a gate voltage. This is a result of the unique electronic structure of BMS, where the valence and conduction bands possess opposite spin polarization when approaching the Fermi level. Our band structure and spin-polarized electronic transport calculations on semi-hydrogenated single-walled carbon nanotubes confirm the existence of BMS materials and demonstrate the electrical control of spin-polarization in them.

Incorporating the spin degree of freedom of electrons into conventional charge-based electronics or using the spin alone leads to the new idea of spintronics, which holds the advantages of low energy consumption and high speed [1-7]. To obtain high-performance spintronic devices, there are still several challenges ahead, including spin-polarized carrier injection, long-distance spin-polarized transport, and effective manipulation and detection of the carriers' spin orientation [5-7].

Half-metals (HMs) are an important kind of spintronics materials with one metallic spin channel and another semiconducting channel. They can thus provide completely spin-polarized current [8,9]. Generally, external magnetic field is required to switch the spin polarization direction. Electric field can be easily generated locally on-chip, and it is thus more convenient in electronics. Creating, manipulating, and detecting magnetic properties via electrical means have been pursued for a long time [10-16]. However, an electric field induced flip of the spin polarization of a 100% polarized current is still challenging.

In this letter, we propose a new class of materials called bipolar magnetic semiconductor (BMS), which can provide completely spin-polarized currents with tunable spin polarization simply by applying a gate voltage. Such a controllability of the spin polarization of current opens new avenues for spintronic devices.

A schematic plot of the density of states of BMS is shown in figure 1(a), together with those of general magnetic semiconductor [figure 1(b)], HM [figure 1(c)], and spin gapless semiconductor [17] [figure 1(d)] for comparison. For BMS, the valence bands (VB) and conduction bands (CB) approach the Fermi level through opposite spin channels. A typical BMS can be described by three energy parameters (Δ1, Δ2, and Δ3), as shown in figure 1(a). Δ1 represents the spin-flip gap between VB and CB edges from different spin channels. Δ1+Δ2 and Δ1+Δ3 reflect the spin-conserved gaps for two spin channels, respectively.

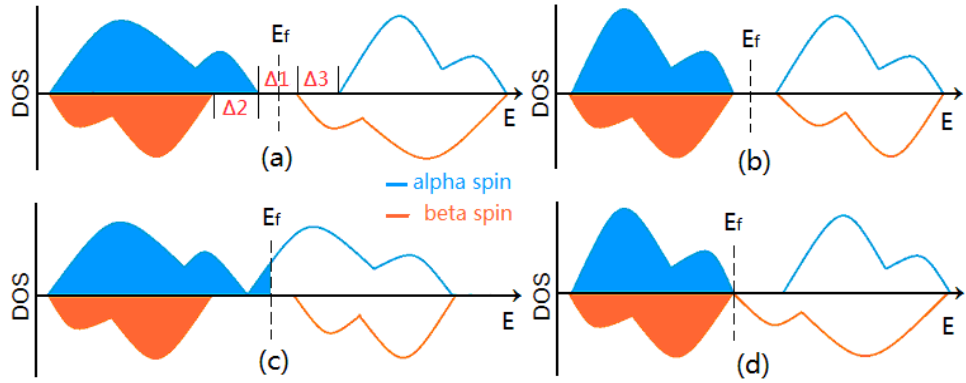

**Figure 1** Schematic density of states of (a) bipolar magnetic semiconductor, (b) general magnetic semiconductor, (c) half metal, and (d) spin gapless semiconductor.

The unique electronic structure of BMS enables a feasible approach to realize half-metallicity in BMS simply by adjusting the position of Fermi level. In particular, since VB and CB approach the Fermi level through opposite spin channels, the half-metallicity in BMS has reversible spin polarization. It can be either alpha spin direction or beta spin direction polarized, depending on whether the Fermi level crossings the VB or CB, respectively. The manipulation of the spin polarization of currents through BMS materials by altering the sign of gate voltage instead of external magnetic field opens up a new opportunity to fabricate electrically-controlled spintronic devices. For practical applications, Δ1 should have a small value to ensure the feasibility to manipulate the spin polarization of currents through BMS by a moderate shift of the Fermi level. When Δ1 equals to zero, BMS becomes a special type of spin gapless semiconductor [17]. At the same time, Δ2 and Δ3 should be large enough to enhance the stability of half-metallicity in BMS as the Fermi level crossings VB or CB.

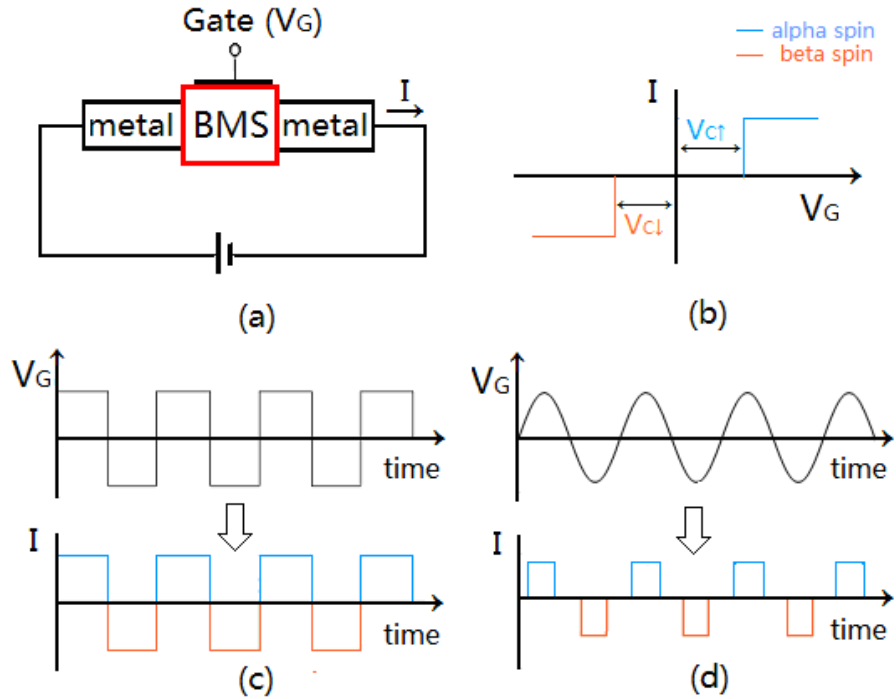

**Figure 2** Schematic plot of BFESF and its I-$V_G$ relationships. (a) Structural model of BFESF. I-$V_G$ relationships of BFESF working under (b) static gate voltage, or oscillated gate voltage in (c) rectangular and (d) sinusoidal forms, respectively.

One of the simplest BMS based spintronic devices is a bipolar field-effect spin-filter (BFESF), as illustrated in figure 2(a). In a BFESF device, BMS is sandwiched between two metallic leads both of which should have long spin-scattering distance. A gate voltage ($V_G$) is applied on BMS to tune the position of Fermi level. As the Fermi level is shifted down into the Δ2 energy window, the electron transport channel becomes alpha spin direction polarized, and it is beta spin direction polarized when the Fermi level moves up into the Δ3 energy window. Fig.

2(b) schematically shows the relationship between spin-polarized current and applied gate voltage. The positive and negative values of current denote the alpha spin and beta spin polarization, respectively. In figure 2(b), the currents transport in two opposite spin channels according to the sign of the $V_G$ (i.e. alpha spin at $V_G > 0$ and beta spin at $V_G < 0$). The threshold voltage ($V_C$) is determined by the spin-flip gap ($\Delta 1$).

Also, when an oscillated gate voltage in rectangular or sinusoidal pulses is applied and the maximum value of $V_G$ is larger than $V_C$ (the upper panels in figure 2(c) and (d)), the output current presents an alternative alpha-spin and beta-spin polarization with the same frequency as gate voltage (the lower parts in figure 2(c) and (d)). In this sense, BMS based BFESF can work as a current modulator, e.g. field-effect spin-rectifier (FESR), which provides a spin impulsive signal of current for spintronic applications.

These two simple cases clearly demonstrate the electrical control of creation and manipulation of spin-polarized currents through BMS materials simply by applying a gate voltage. The next question is whether BMS materials exist. Based on the first-principles density functional theory (DFT) method, we show that semi-hydrogenated SWCNTs are good candidates for BMS materials. Carbon nanomaterials, including graphene and SWCNTs, are attractive for spintronic applications due to their long spin-scattering lengths [18-21]. Using non-equilibrium Green's function (NEGF) method, we also directly demonstrate that completely spin-polarized currents with reversible spin-polarization can be realized in BMS materials by applying a gate voltage. A BFESF device model is fabricated by using one single SWCNT, where a semi-hydrogenated SWCNT segment is sandwiched between two pristine SWCNT electrodes.

The electronic structures of semi-hydrogenated SWCNT are calculated based on DFT method within the Perdew-Burke-Ernzerholf generalized gradient approximation (GGA) [22] implemented in Vienna ab initio simulation package (VASP) [23]. The projector augmented wave (PAW) potential and the plane-wave cut-off energy of 400 eV are used. The first Brillouin zone is sampled with Monkhorst-Pack grid of $1\times1\times24$ and $1\times1\times12$ for armchair and zigzag-type SWCNT, respectively. Both the lattice constant along the axial direction and the positions of all atoms are relaxed until the force is less than 0.01 eV/Å. The criterion for the total energy is set as $1\times10^{-5}$ eV. The nearest distance between two neighboring SWCNTs is about 10 Å. To investigate the magnetic ordering in semi-hydrogenated SWCNTs, a supercell with double periodic units along the tube direction is used for armchair type SWCNT. For the zigzag type SWCNT, the unit cell is used. Since the GGA approach usually underestimates the band gap of semiconductor, the HSE06 screened hybrid functional theory method [24,25] is adopted to check the calculated electronic band structures of armchair type semi-hydrogenated (5,5) SWCNT.

The structure of semi-hydrogenated SWCNT based molecular junction is relaxed within the local spin density approximation and double-ζ plus polarization basis set is used, implemented in SIESTA package [26]. The energy cutoff for real-space mesh size is set to be 300 Ry and the force tolerance is 0.01eV/Å. The electronic transport calculations are performed by using real-space NEGF techniques implemented in the ATK package [27,28]. To save the computational consumption, LSDA functional with a single-ζ basis set and a mesh cutoff of 150 Ry was used. These settings have been tested valid for carbon based materials [29]. In ATK, the gate electrode is not included as a physical electrode and it is assumed to induce an external potential localized at the molecular region by simply shifting the molecular projected self-consistent Hamiltonian (MPSH) part.

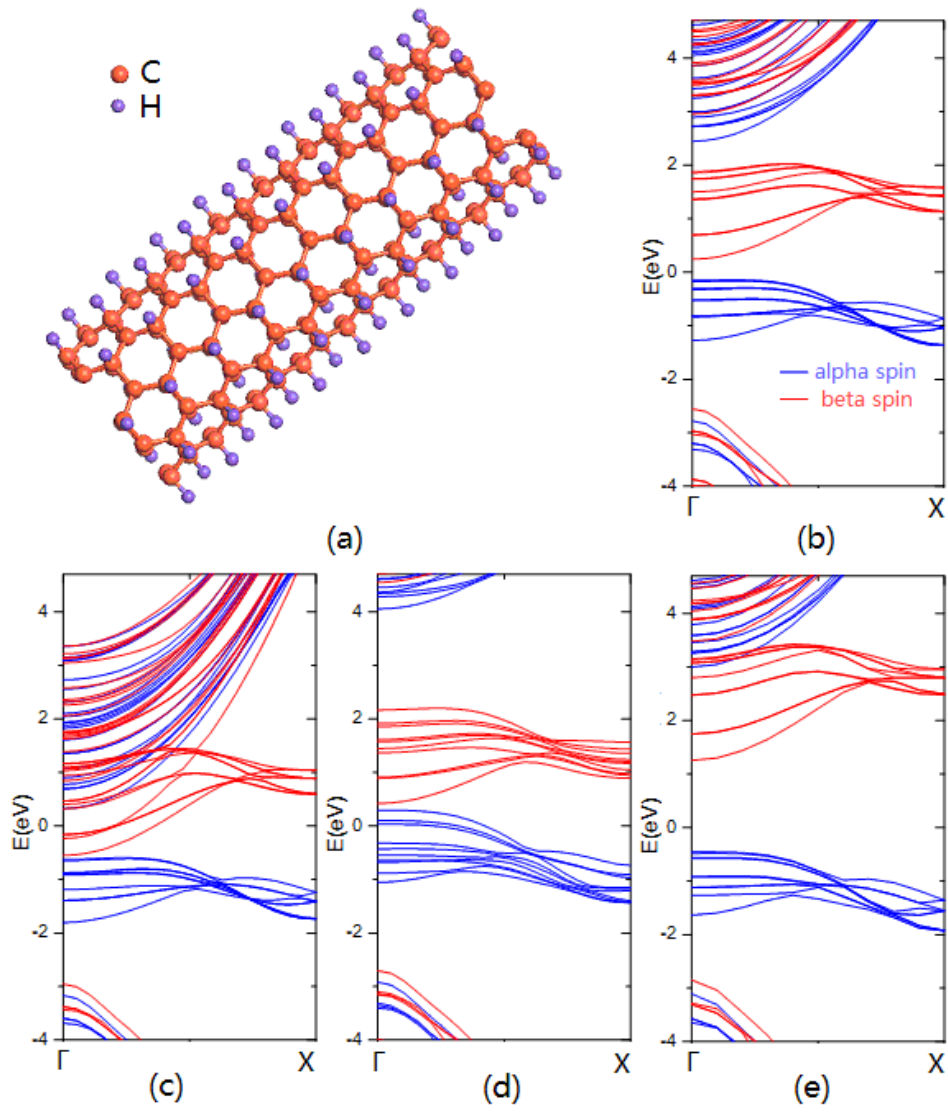

**Figure 3** (a) Optimized structure of semi-hydrogenated (5,5) SWCNT with PBE functional. (b) Electronic band structure of semi-hydrogenated (5,5) SWCNT. (c) Electronic band structure of electron-doped semi-hydrogenated (5,5) SWCNT with a doping concentration of 0.03 electron per atom. (d) Electronic band structure of hole-doped semi-hydrogenated (5,5) SWCNT with a doping concentration of 0.03 hole per atom. (e) Electronic band structure of semi-hydrogenated (5,5) SWCNT based on HSE06 calculations. Purple and red balls represent hydrogen and carbon atoms, respectively. The Fermi level is set at zero.

In SWCNT, carbon atoms are arranged in two sublattices and the $sp^2$ hybridization of $2s$ and two $2p$ carbon orbitals leads to σ bonding of carbon atoms within the tube wall. The third $2p$ orbital ($p_z$) perpendicular to the tube wall contributes to delocalized π bonds. In semi-hydrogenated SWCNTs, only carbon atoms from one of the two sublattices are hydrogenated. The optimized structure of a semi-hydrogenated (5,5) SWCNT is shown in figure 3(a). Semi-hydrogenation of SWCNT saturates the $p_z$ orbitals of half of the carbon atoms, thus destroys the extensive *π-type* bonds. The left unpaired $p_z$ electrons on the other half of carbon atoms are spin-polarized. The calculated total magnetism for semi-hydrogenated (5,5) SWCNT is 10 $\mu_B$ per unit cell, which contains 10 hydrogen atoms, 10 hydrogenated and 10 unhydrogenated carbon atoms. The Bader charge analysis results show that there is a localized moment of about 0.81 $\mu_B$ on each magnetic carbon atom and about 0.17 $\mu_B$ on hydrogen atom, as shown in figure 4(a).

To determine the magnetic ground state of semi-hydrogenated SWCNTs, the ferromagnetic (FM) order (figure 4(b)) and two antiferromagnetic orders (AFM1 and AFM2 in figure 4(c) and (d)) are studied. In AFM1 state, the magnetic carbon atoms are in FM ordering along the tube's axial direction and AFM ordering along the circumference direction (figure 4(c)).  In AFM2 state, the magnetic carbon atoms are in AFM ordering along the tube's axial direction, whereas FM ordering along the circumference direction (figure 4(d)). The energy difference between FM and AFM2 states for both zigzag and armchair-type semi-hydrogenated SWNCTs with different diameters are summarized in Table I. The AFM2 state is energetically more favorable than AFM1 state. Except zigzag-type semi-hydrogenated SWCNTs with small dimater, *e.g.* (8,0) and (10,0) semi-hydrogenated SWCNT, all investigated tubes have FM ground states. The energy difference between FM and AFM states ranges from 0.491 to 1.961 *eV* per supercell, where the large-diameter semi-hydrogenated SWCNT has large FM-AFM energy difference. The nonmagnetic (NM) states of semi-hydrogenated SWCNT are also considered, which are remarkably unstable than their FM states. For example, the NM state of semi-hydrogenated (5,5) SWCNT is less stable than its' FM with an energy difference of about 5.33 *eV* per supercell.

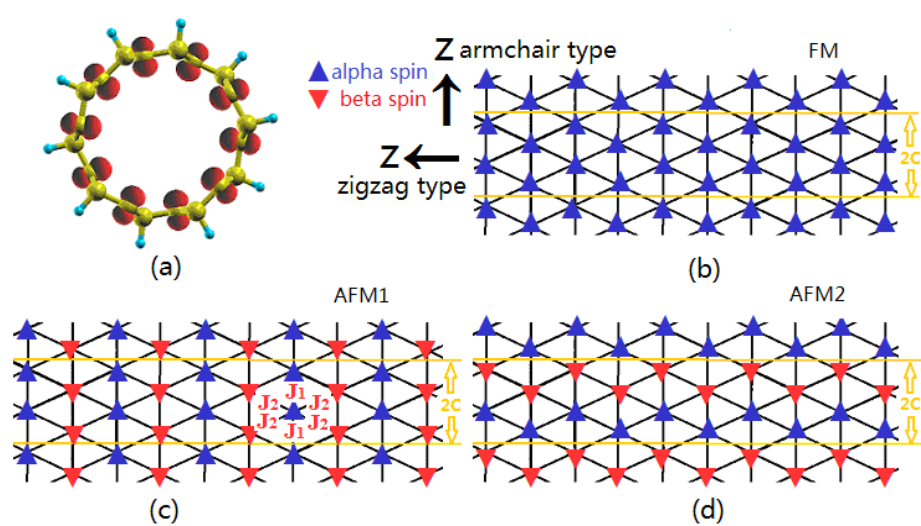

**Figure 4** (a) Isosurface of spin density for ferromagnetic semi-hydrogenated (5,5) SWCNT with isovalue 0.22 $e/Å^3$. (b)-(d) FM, AFM1 and AFM2 states. The nanotube is tailored along its axis i.e. Z direction and spread as a sheet, where the hydrogen atoms and the hydrogenated carbon atoms carry little magnetic moments and are not shown.

It is known that one-dimensional long-range magnetic order is prohibited at a finite temperature based on the spin lattice models [30]. However, taking into account effects of kinetic barriers to reaching equilibrium or the existence of substrate, short-range magnetic order at nanometer-scale is still possible [31-33].

**Table:** The calculated lattice constants, energy differences between FM and AFM states, effective exchange parameters for both armchair and zigzag semi-hydrogenated SWCNTs with various diameters are summarized.

| (n,m) | (5,5) | (10,10) | (8,0) | (10,0) | (12,0) | (14,0) | (18,0) |
|---|---|---|---|---|---|---|---|
| Lattice constant C (Å) | 2.53 | 2.53 | 4.35 | 4.37 | 4.37 | 4.37 | 4.38 |
| E(AFM1)-E(FM) (eV) | 0.774 | 1.961 | 0.453 | 0.796 | 1.057 | 1.265 | 1.623 |
| E(AFM2)-E(FM) (eV) | 0.491 | 1.445 | -0.665 | 0.045 | 0.505 | 0.821 | 1.263 |
| $J_1$ (meV) | 2.6 | 5.8 | -27.8 | -8.8 | 0.5 | 3.4 | 6.3 |
| $J_2$ (meV) | 9.9 | 12.3 | 7.1 | 9.9 | 11.0 | 11.3 | 11.3 |

The band structure of semi-hydrogenated (5,5) SWCNT at ferromagnetic ground state indicates that it is a ideal BMS materials (figure 3(b)). As illustrated in figure 3(c) and (d), electron or hole-doping in semi-hydrogenated (5,5) SWCNT introduces robust half-metallicity with opposite spin polarizations. We note that density functional used in this study may underestimate the band gap Δ1 [34]. Our test calculations with screened hybrid functional indeed give an enlarged spin-flip gap, but the overall band structure is similar (figure 3(e)). We also perform test calculations on the semi-hydrogenated SWCNTs with different diameters and chirality. The results show that all semi-hydrogenated armchair SWCNTs are robust BMS materials and those semi-hydrogenated zigzag SWCNTs with large diameter, such as (14,0) zigzag SWCNTs, are BMS materials too.

Based on nearest-neighbor Heisenberg model, the effective Heisenberg Hamiltonian can be written as $H_{eff}= -\sum J_{ij} \cdot e_i \cdot e_j$, where $e_i$ is the unit vector pointing in the direction of the magnetic moment at site $i$, $J_{ij}$ is effective exchange parameter. In semi-hydrogenated SWCNT, there are two types of $J$-coupling in this system ($J_1$ and $J_2$), which measure the exchange interaction between magnetic carbon atoms in the circumference and axis directions (figure 4(c)), respectively. To simplify the calculation, the angle between $e_i$ and $e_j$ is set to 0 (FM order) or 180 (AFM order) degrees. The calculated $J_1$ and $J_2$ values are 9.7 and 2.6 *meV* for semi-hydrogenated (5,5) SWCNT, comparable with those (range from 10 to 100 meV) in transition metal alloys [35,36].

In this study, we don't consider the spin-orbital interaction in carbon atoms. The small magnitude of spin-orbital interaction in carbon atoms (about 4 meV) should not change the ferromagnetic ground state character of the semi-hydrogenated SWCNT [37-39]. However, the spin-orbital interaction in carbon atoms will determine the spatial direction of spin-polarization in the semi-hydrogenated SWCNT at low-temperature when an external magnetic field is applied.

Moreover, to check the effect of defects on the ability of tuning the Fermi level in semi-hydrogenated SWCNTs, some test calculations for boron and nitrogen substitution defects are performed. In a supercell consist of four repeated units in tube axis direction, one carbon atom with or without hydrogenation is substituted with boron or nitrogen atom. As shown in Fig. 5, the defect-induced impurity states are deep in energy, which will thus not lead to the Fermi-level trap problem in semi-hydrogenated SWCNTs.

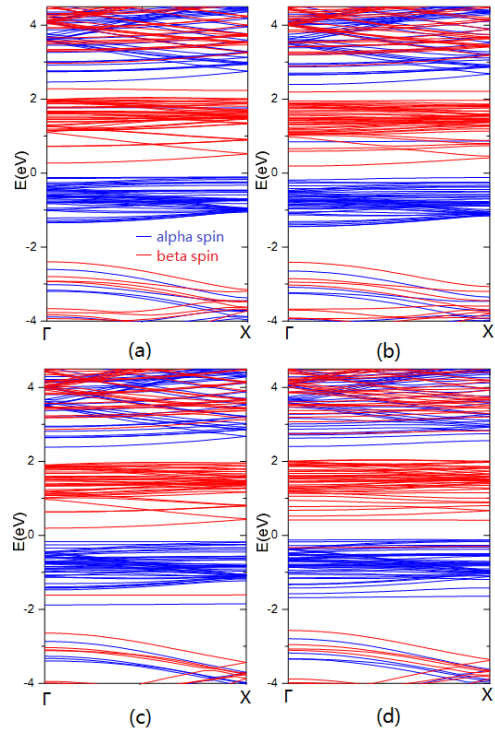

**Figure 5.** The electronic band structures of semi-hydrogenated (5,5) SWCNT with defect: (a) boron substitution of an unhydrogenated carbon atom, (b) boron substitution of a hydrogenated carbon atom, (c) nitrogen substitution of an unhydrogenated carbon atom, and (d) nitrogen substitution of a hydrogenated carbon atom. The Fermi level is set as zero.

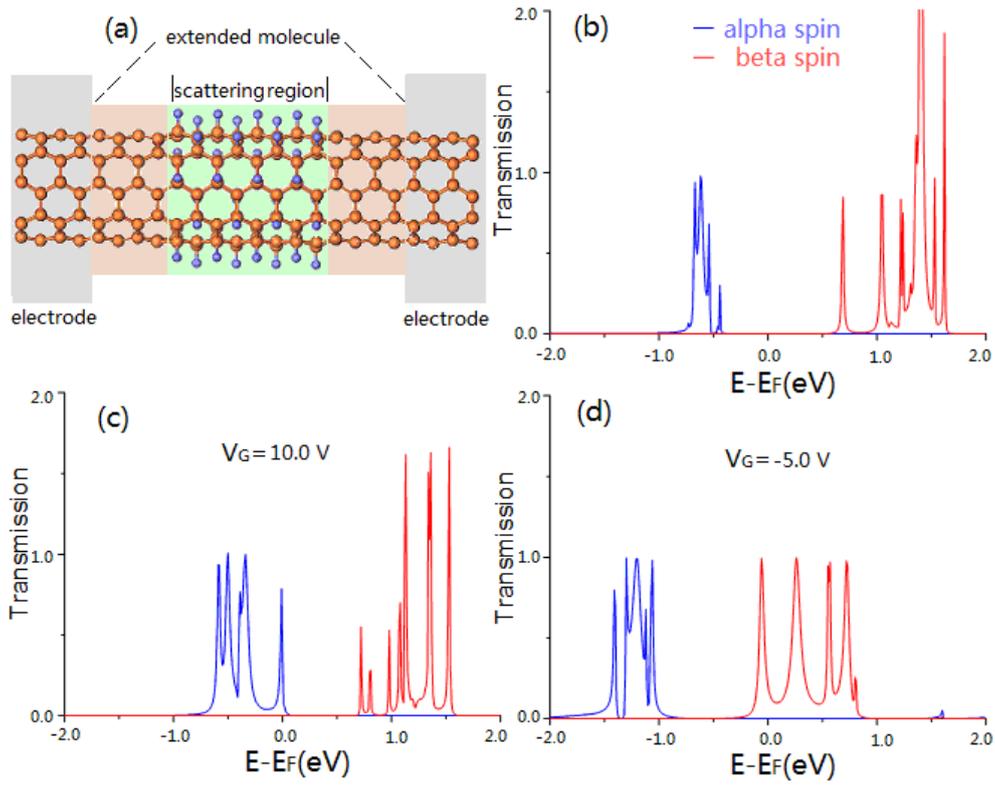

**Figure 6** (a) Model of the semi-hydrogenated (5,5) SWCNT based BFESF device. (b) Transmission spectrum at zero bias without a gate voltage. (c) Transmission spectrum at zero bias with a gate voltage of 10.0 V.  (d) Transmission spectrum at zero bias with a gate voltage of -5.0 V.

Based on the semi-hydrogenated SWCNT, a BFESF device can be constructed by sandwiching it between two pristine SWCNT electrodes, as shown in figure 6(a). The device model is divided into three parts, including left electrode, extended molecule, and right electrode. Both electrodes are semi-infinite metallic (5,5) SWCNTs. The extended molecule consists of a scattering region of four repeated units of semi-hydrognated (5,5) SWCNT segment and two unit cells of (5,5) SWCNTs in both sides.

The calculated transmission spectrum $T(E)$ of the device at zero bias without a gate voltage is shown in figure 6(b). Transmission peaks belonging to two opposite spin channels can be distinguished below and above the Fermi level, respectively. This result suggests that the BMS characteristics of semi-hydrogenated (5,5) SWCNT are well conserved after it is sandwiched between two SWCNT electrodes. When applying a gate voltage of 10.0 V, the Fermi energy level is shifted downward and the transmission peak for alpha-spin channel dominates the transmission spectrum at the Fermi level (figure 6(c)). The spin polarization, defined as the absolute value of [T(alpha-spin) - T(beta-spin)] / [T(alpha-spin) + T(beta-spin)] at zero bias, reaches 99.9%. When applying a gate voltage of -5.0 V, the Fermi level moves up and the transmission peak for beta-spin channel dominates the transmission spectrum at the Fermi level with a 100% spin-polarization, as shown in figure 6(d). Thus, the device can work at two modes with opposite spin polarization tuned by the gate voltage. The unsymmetrical response of bands shift to positive and negative gate voltage comes from the different spatial distributions of the highest occupied and lowest unoccupied states.

BMS materials provide a feasible solution for manipulation of spin-polarized currents and hold many advantages in spintronic applications. Although electrical means have been proved successfully to alter the magnetic properties in many materials, BMS materials present for the first time the opportunity to provide completely spin-polarized carriers with reversible spin polarization tuned by electron or hole doping. The electrical control of spin polarization of carriers in BMS originates from their unique electronic structure and can be simply realized by adjusting the Fermi level, which is quite different from those means by using spin-orbit coupling of electron. Thus, the manipulation of spin polarization of carriers in BMS is more convenient by using mature techniques in the semiconductor industry.

Also, as one type of magnetic semiconductors, current applications of magnetic semiconductors can be directly extended to BMS materials. Moreover, the controllability of the spin-polarization with gate-voltage makes BMS a more versatile material. For example, it is possible to simultaneously apply two gates on BMS to construct different spin-polarization transport channels in a single material.

Actually, the means to manipulate the spin-polarization in BMS materials are not limited to electric field. Some other physical or chemical methods, such as doping, applying strain, or chemical functionalization *etc.*, may be also feasible to reach the goal discussed above. Therefore, using these methods alone or combining them with external magnetic field should open up great opportunities to enrich the family of spintronic devices with multiple functions. Moreover, using different electrodes, such as ferromagnetic materials, superconductors, or topological insulators, may also lead to novel spin transport behavior different from those with metallic electrodes.

BMS may be explored also in other materials from zero to three dimensions. Some single molecular magnets and doped semiconductors may be good candidates for BMS materials. The above discussed electrical control of spin polarization can also be extended to other types of materials with similar band structures characteristics as BMS. For example, a bipolar HM, of which the band structure is similar as that of BMS with different position of Fermi level, can also provide completely spin-polarized currents with reversible spin polarization by applying a gate voltage or not.

In conclusion, we have proposed a new class of materials, called BMS, which can provide completely spin-polarized currents with reversible spin-polarization simply by applying a gate voltage. The concept of BMS is verified by first-principles electronic and transport calculations on semi-hydrogenated SWCNTs. BMS materials will certainly enrich the family of spintronic devices. The combination of electrical control with other physical means, such as magnetic field, is expected to make BMS based multifunctional spintronic device possible. This new class of materials thus gives us a great flexibility in spintronic applications.

**Acknowledgements**

This work is partially supported by the National Key Basic Research Program (2011CB921404, 2012CB922001), by NSFC (21121003, 91021004, 20933006, 11004180, 51172223), and by USTCSCC, SCCAS, Tianjin, and Shanghai Supercomputer Centers.

*Email: jlyang@ustc.edu.cn

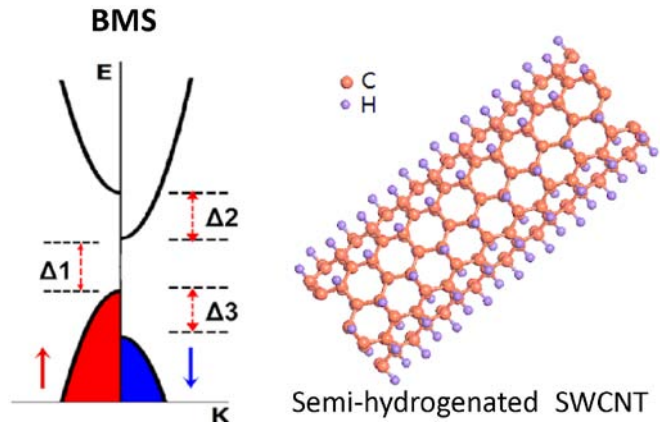

A new concept of bipolar magnetic semiconductor (BMS) with gate-controlled spin-polarization is proposed and confirmed in semi-hydrogenated SWCNT.